# Software (Re-)Engineering with PSF III: an IDE for PSF

*Bob Diertens*

Programming Research Group, Faculty of Science, University of Amsterdam

*ABSTRACT*

We describe the design of an integrated development environment (IDE) for PSF. In the software engineering process we used process algebra in the form of PSF for the specification of the architecture of the IDE. This specification is refined to a PSF specification of the IDE system as a ToolBus application, by applying vertical and horizontal implementation techniques. We implemented the various tools as specified and connected them with a ToolBus script extracted from the system specification.

**1. Introduction**

We developed an integrated development environment (IDE) for PSF (Process Specification Formalism) with the use of PSF in the software engineering process. Use of PSF in the development of software systems has been investigated in Diertens [5] and [6]. This resulted in two PSF libraries, one library providing primitives and environment for specification of the architecture of the software system and one providing primitives and environment for specifying the system as a ToolBus application. To get a ToolBus application specification from the architecture specification the refining techniques horizontal implementation and vertical implementation are applied.

Development of large software systems is a difficult task. We use process algebra in the design phase of the software systems to make the software design process easier. The use of process algebra is not to alter the design process, but to support it whatever the design process may be. A strong argument for using process algebra in the software design process is the possibility to validate the design by simulation and animation.

Previous papers in our series on software (re-)engineering with PSF, focused on the redesign of existing tools in the PSF Toolkit. This paper describes the development of a new tool, and should found our confidence in using PSF and the PSF Toolkit in the software design process. It must be noted that, although presented here straightforward, the design of the IDE was an iterative process.

PSF is based on ACP (Algebra of Communicating Processes) [1] and ASF (Algebraic Specification Formalism) [2]. A description of PSF can be found in [14], [15], [7], and [8]. The PSF Toolkit contains among other components a compiler and a simulator that can be coupled to an animation [9]. Animations can either be made by hand or be automatically generated from a PSF specification [10]. The animations play an important role in our software development process as they can be used to test the specifications and are very convenient in communication to other stakeholders.

The ToolBus [3] is a coordination architecture for software applications developed at the CWI (Amsterdam) and the University of Amsterdam. It utilizes a scripting language based on process algebra to describe the communication between software tools. A ToolBus script describes a number of processes that can communicate with each other and with various tools existing outside the ToolBus. The role of the ToolBus when executing the script is to coordinate the various tools in order to perform some complex task.

We start the description of the design of the IDE for PSF with the requirements in section 2, followed by the specification of the architecure in section 3. In section 4 the architecture is refined to obtain a specification of the system as a ToolBus application. And the implementation of the IDE is described in section 5.

## 2. Requirements for the IDE

Every user of the PSF Toolkit has its own preferences in the way the tools are applied. Some prefer to be in full control and use a command line approach, or automate the execution of the tools with script or make-like facilities. Others prefers the integration of the tools into one big tool which automates the execution and provides control through a graphical user interface. The purpose of the IDE for PSF is to support the last group.

**Functional Requirements**

- Integration of editing facilities, compiler, simulator, and animation generator.
- Simple interaction with the user through a consistent graphical user interface.
- Providing clear information on the status of the development process.
- Hiding of the interaction between the tools.

**Non-functional Requirements**

- Modular design with easy to replace components.
- Use of existing tools in the PSF Toolkit. Any modifications necessary for the interaction between the tools should be as small as possible and may not alter the command-line interface of the tools.
- Extendable with other tools.

## 3. Architecture Specification of the IDE

We specify software architecture in PSF with the use of a PSF library providing architecture primitives. The primitives are `snd` and `rec` actions for communication, each taking a connection and a data term as argument. A connection can be build up with a connection function >> with two identifiers indicating a component as argument. Processes describing the software architecture with these primitives can be set in an architecture environment, also provided by the PSF library. The architecture environment takes care of encapsulation to enforce the communication between the processes.

To develope a specification of the architecture for the IDE, we start with a simple scenario and try to specify an architecture for just this scenario. We adapt the specification step by step to incorporate other scenario's.

*3.1 Scenario: one module specification*

In this scenario our specification consist of only one module. The module can be edited and compiled until the IDE is stopped.

We need four components, a component for functions on the specification, an editor, a compiler, and a viewer for possible errors from compilation of the specification. We first specify the component identifiers and the data we use in our architecture.

```
data module IDEData
begin
    exports
    begin
        functions
```

```
          function : → ID
          editor : → ID
          compiler : → ID
          errorviewer : → ID

          edit-module : → DATA
          close-module : → DATA
          module-closed : → DATA
          module-written : → DATA
          compile : → DATA
          errors : → DATA
          no-errors : → DATA
      end
   imports
      ArchitectureTypes
end IDEData
```

The functions that can be invoked are to edit, close, and compile the module, and to quit the IDE. We specify the behavior of this component as follows

```
process module Function
begin
   exports
   begin
      processes
         Function
   end
   imports
      IDEData,
      ArchitecturePrimitives
   atoms
      edit-module
      close-module
      compile
      push-quit
   definitions
      Function =
         (
            edit-module .
            snd(function >> editor, edit-module)
         +  close-module .
            snd(function >> editor, close-module)
         +  rec(editor >> function, module-closed)
         +  rec(editor >> function, module-written)
         +  compile .
            snd(function >> compiler, compile) . (
               rec(compiler >> function, errors)
            +  rec(compiler >> function, no-errors)
            )
         ) *
         push-quit .
         snd-quit
end Function
```

The editor can receive requests for starting and closing of an editing session, and has user actions for writing and closing the module. We specify the Editor component as below. In the remainder we restrict ourselves to show the process definition only and leaving out other sections and the modular structure unless we think its necessary for better understanding of the specification.

```
Editor =
   rec(function >> editor, edit-module) .
   start-editor .
   Edit
Edit =
      rec(function >> editor, close-module) .
      close-editor .
      Editor
   +  editor-close .
      snd(editor >> function, module-closed) .
      Editor
   +  editor-write .
      snd(editor >> function, module-written) .
```

```
                    Edit
```

The Compiler component reports either succesfull compilation or unsuccesfull compilation and with the latter also reports the errors encountered.

```
            Compiler =
                rec(function >> compiler, compile) . (
                   snd(compiler >> function, errors) .
                   snd(compiler >> errorviewer, errors)
                +  snd(compiler >> function, no-errors)
                ) * delta
```

The ErrorViewer component just displays the errors from compilation.

```
            ErrorViewer =
                (
                    rec(compiler >> errorviewer, errors)
                ) * delta
```

We specify the system consisting of the components in parallel as follows.

```
        process module IDESystem
        begin
            exports
            begin
                processes
                    IDESystem
            end
            imports
                Function,
                Editor,
                Compiler,
                ErrorViewer
            definitions
                IDESystem =
                      Function
                   ‖  Editor
                   ‖  Compiler
                   ‖  ErrorViewer
        end IDESystem
```

We put this system in the architecture environment by means of binding the main process to the System parameter of the Architecture module from the Architecture library.

```
        process module IDE
        begin
            imports
                Architecture {
                    System bound by [
                        System → IDESystem
                    ] to IDESystem
                    renamed by [
                        Architecture → IDE
                    ]
                }
        end IDE
```

The generated animation of this system is shown in Figure 1.

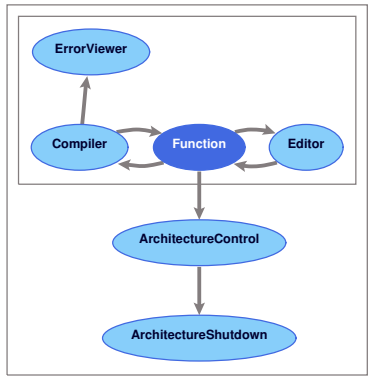

**Figure 1.** Animation of architecture for single module specifications

*3.2 Scenario: multiple module specification*

In the next scenario, we deal with a specification consisting of more than one module. To manage the modules we split the Function component into a new Function component with only a quit action and a module manager.

```
Function =
   push-quit .
   snd(function >> module-manager, quit)
```

The quit action is send to the module manager, which can decide on what actions to take before the actual quitting of the system.

```
ModuleManager =
   (
      edit-module .
      (
         EventsEditorManager *
         snd(module-manager >> editor-manager, edit-module)
      )
   + close-module .
      (
         EventsEditorManager *
         snd(module-manager >> editor-manager, close-module)
      )
   + EventsEditorManager
   + compile .
      snd(module-manager >> compiler, compile) . (
         rec(compiler >> module-manager, errors)
      + rec(compiler >> module-manager, no-errors)
      )
   + rec(function >> module-manager, quit) .
      snd-quit
   ) * delta
EventsEditorManager =
      rec(editor-manager >> module-manager, module-closed)
   + rec(editor-manager >> module-manager, module-written)
```

We use the construction with the process EventsEditorManager to prevent deadlocks, which otherwise can occur when the module manager and the editor manager want to send messages to each other at the same time.

We also introduce an editor manager for dealing with multiple editors.

```
EditorManager =
   (
      rec(module-manager >> editor-manager, edit-module) .
      start-editor
   + editor-close .
      snd(editor-manager >> module-manager, module-closed)
```

```
          +  editor-write .
             snd(editor-manager >> module-manager, module-written)
          +  rec(module-manager >> editor-manager, close-module) .
             close-editor
       ) * delta
```

At any time, the module manager can request to start or to close an editor for a module. An editor is expected to report on a closure and on writing of the module. We do not keep track on how many editors there are open at a certain moment, that is up to the implementation of the editor manager.

Our changes of the architecture to accommodate the scenario results in the generated animation as shown in Figure 2.

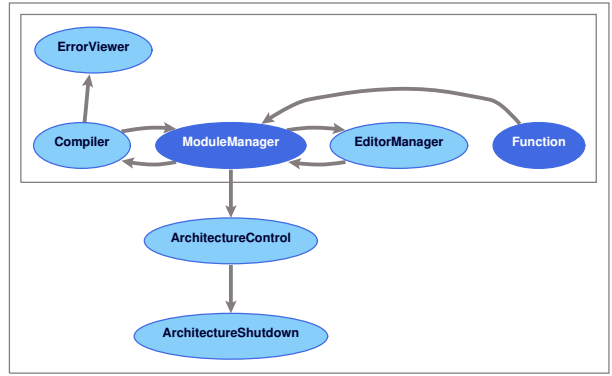

**Figure 2.** Animation of architecture for multi module specifications

*3.3 Scenario: partial compilation*

With the previous scenario, we compile the complete specification everytime. The PSF compiler however, only applies its steps (parsing, normalizing, flattening) if necessary for a module, based on the timestamps of the PSF module and the intermediate files. We want to make it possible for the module manager to issue parse, compile, and flatten request whenever it wants. At the same time, we not only want the compiler to respond to the request of the module manager, but also act on its own. This scheme will be restricted by the implementation of the module manager and compiler (see section 4.2, page 11).

We alter the module manager and the compiler components to provide the described behavior.

```
       ModuleManager =
          .
          .
          .
       +  parse .
          (
             EventsCompiler *
             snd(module-manager >> compiler, parse)
          )
       +  compile .
          (
             EventsCompiler *
             snd(module-manager >> compiler, compile)
          )
       +  flatten .
          (
             EventsCompiler *
             snd(module-manager >> compiler, flatten)
          )
       +  EventsCompiler
          .
          .
          .
```

We replaced the previous compile action with separate parse, compile, and flatten actions. In the same manner as with events from the editor manager, we make a construction for dealing with events from the compiler in order to prevent deadlocks.

```
EventsCompiler =
      rec(compiler >> module-manager, parse-ok)
    + rec(compiler >> module-manager, parse-uptodate)
    + rec(compiler >> module-manager, parse-error)
    + rec(compiler >> module-manager, compile-ok)
    + rec(compiler >> module-manager, compile-uptodate)
    + rec(compiler >> module-manager, compile-error)
    + rec(compiler >> module-manager, flatten-ok)
    + rec(compiler >> module-manager, flatten-uptodate)
    + rec(compiler >> module-manager, flatten-error)
```

The compiler receives parse, compile, and flatten requests from the module manager and sends results of parse, compile, and flatten action to the module manager.

```
Compiler =
    (
      rec(module-manager >> compiler, parse)
    + parse-ok .
      snd(compiler >> module-manager, parse-ok)
    + parse-uptodate .
      snd(compiler >> module-manager, parse-uptodate)
    + parse-error .
      snd(compiler >> module-manager, parse-error) .
      snd(compiler >> errorviewer, errors)
    + rec(module-manager >> compiler, compile)
    + compile-ok .
      snd(compiler >> module-manager, compile-ok)
    + compile-uptodate .
      snd(compiler >> module-manager, compile-uptodate)
    + compile-error .
      snd(compiler >> module-manager, compile-error) .
      snd(compiler >> errorviewer, errors)
    + rec(module-manager >> compiler, flatten)
    + flatten-ok .
      snd(compiler >> module-manager, flatten-ok)
    + flatten-uptodate .
      snd(compiler >> module-manager, flatten-uptodate)
    + flatten-error .
      snd(compiler >> module-manager, flatten-error) .
      snd(compiler >> errorviewer, errors)
    ) * delta
```

As mentioned above, we want the compiler to honour the requests of the module manager and also the possibility to act on its own. Therefore, we did not specify a relation between a request and sending the result of an action here.

*3.4 Scenario: import modules from a library*

Some of the modules which are imported may come from a library. We introduce a library manager for adding, deleting and re-order a list of libraries. Every time a change in of this list occurs the list has to be send to the compiler, so that it knows were to look for imported modules.

We specify the library manager as follows.

```
LibraryManager =
    (
       set-libraries .
       snd(library-manager >> compiler, set-libraries)
    ) * delta
```

To the specification of the compiler we add the following alternative.

```
    + rec(library-manager >> compiler, set-libraries)
```

*3.5 Scenario: simulation*

Next to compilation, an IDE should also support tools to act on the compiled specification, such as a simulator. These tools can act on a TIL (Tool Interface Language) specification. A TIL specification is the result of compilation of the a PSF specification.

We add to the module manager the alternatives to send new and delete notifications on TIL specifications.

```
+   new-tilspecification .
    snd(module-manager >> simulator, new-tilspecification)
+   delete-tilspecification .
    snd(module-manager >> simulator, delete-tilspecification)
```

We add a simulator component to our specification.

```
Simulator =
  (
      rec(module-manager >> simulator, new-tilspecification)
  +   rec(module-manager >> simulator, delete-tilspecification)
  ) * delta
```

The actual running of a simulator is an internal action of this component that we will specify on a lower level of our design.

The resulting animation is shown in Figure 3.

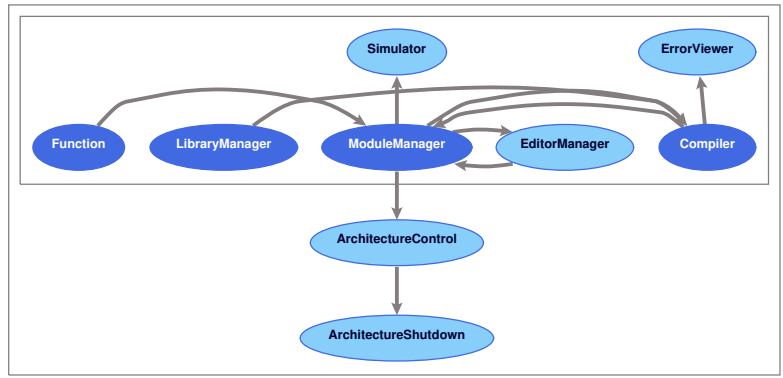

**Figure 3.** Animation of architecture with simulator[1]

*3.6 Scenario: simulation and animation*

Simulation can be run together with an animation. Such an animation can be generated from a TIL specification by the animation generator of the PSF Toolkit. To use the animation generator in the IDE, the module manager also have to send new and delete notifications to a animation generator component. We specify the animation generator as follows.

```
AnimationGenerator =
  (
      rec(module-manager >> animation-generator, new-tilspecification)
  +   rec(module-manager >> animation-generator, delete-tilspecification)
  +   new-animation .
      snd(animation-generator >> simulator, new-animation)
  +   animationgeneration-error .
      snd(animation-generator >> simulator, animationgeneration-error) .
      snd(animation-generator >> errorviewer, errors)
  ) * delta
```

We also add alternatives to the simulator for receiving the notifications from the animation generator.

---

1. The layout of the components in the animation is generated by the program dot, which is part of the graph visualization software package Graphviz [11]. It is not always the best layout possible.

## 4. System Specification of the IDE

We derive a specification of the IDE as a ToolBus application from the specification of the architecture of the IDE by applying the implementation techniques *action refinement* and *process constraining* described in [6]. Action refinement or *vertical implementation* is a technique for mapping abstract actions onto concrete processes. Process constraining or *horizontal implementation* is a technique in which a process is put in parallel with another process and where communication between these processes is enforced by encapsulation. In recent work by Bergstra and Middelburg [4] this is called component composition of an interface and a behaviour process.

A specification of a ToolBus application consist of specifications for the processes inside the ToolBus, the ToolBus script, and specifications for the tools outside the ToolBus with which the ToolBus processes communicate. There are two sets of primitives, one sets for communications between the processes inside the ToolBus, and one set for communications between the ToolBus processes and the tools. The first set consist of `tb-snd-msg` and `tb-rec-msg` each taking three arguments, the identifier of the sender, the identifier of the receiver, and a term. The second set consists of the ToolBus process actions `tb-rec-event`, `tb-snd-ack-event`, `tb-snd-do`, `tb-snd-eval`, and `tb-rec-value`, taking a tool identifier and a term as arguments, and the tool actions `tooltb-snd-event`, `tooltb-rec-ack-event`, `tooltb-rec`, `tooltb-snd-value`, taking a single term as argument.

*4.1 Action refinement*

We show the refinings for a vertical implementation of the architecture specification. We start with some default mappings, which are to be applied when there are no other mappings to apply.

```
snd($1 >> $2, $3)    → tb-snd-msg($1, $2, tbterm($3))
rec($1 >> $2, $3)    → tb-rec-msg($1, $2, tbterm($3))
```

The $n on the left hand side represent matched terms that have to be filled in on the right hand side. Below the mappings per module (component) are given.

**Function**
```
push-quit           → tb-rec-event(MODULEMANAGER, tbterm(quit)) .
                      tb-snd-ack-event(MODULEMANAGER, tbterm(quit))
```

**ModuleManager**
```
edit-module         → tb-rec-event(MODULEMANAGER, tbterm(edit-module)) .
                      tb-snd-ack-event(MODULEMANAGER, tbterm(edit-module))
close-module        → tb-rec-event(MODULEMANAGER, tbterm(close-module)) .
                      tb-snd-ack-event(MODULEMANAGER, tbterm(close-module))
parse               → tb-rec-event(MODULEMANAGER, tbterm(parse)) .
                      tb-snd-ack-event(MODULEMANAGER, tbterm(parse))
compile             → tb-rec-event(MODULEMANAGER, tbterm(compile)) .
                      tb-snd-ack-event(MODULEMANAGER, tbterm(compile))
flatten             → tb-rec-event(MODULEMANAGER, tbterm(flatten)) .
                      tb-snd-ack-event(MODULEMANAGER, tbterm(flatten))
new-tilspecification →
                      tb-rec-event(MODULEMANAGER,
                          tbterm(new-tilspecification)) .
                      tb-snd-ack-event(MODULEMANAGER,
                          tbterm(new-tilspecification))
delete-tilspecification →
                      tb-rec-event(MODULEMANAGER,
                          tbterm(delete-tilspecification)) .
                      tb-snd-ack-event(MODULEMANAGER,
                          tbterm(delete-tilspecification))
rec(compiler >> module-manager, $1)    →
                      tb-rec-msg(compiler, module-manager, tbterm($1)) .
                      tb-snd-do(MODULEMANAGER, tbterm($1))
rec(editor-manager >> module-manager, $1)   →
                      tb-rec-msg(editor-manager, module-manager,
```

```
                               tbterm($1)) .
                               tb-snd-do(MODULEMANAGER, tbterm($1))
       snd-quit              → snd-tb-shutdown
```

### EditorManager

```
       start-editor          → tb-snd-do(EDITORMANAGER, tbterm(start-editor))
       editor-close          → tb-rec-event(EDITORMANAGER, tbterm(editor-close)) .
                               tb-snd-ack-event(EDITORMANAGER, tbterm(editor-close))
       editor-write          → tb-rec-event(EDITORMANAGER, tbterm(editor-write)) .
                               tb-snd-ack-event(EDITORMANAGER, tbterm(editor-write))
       close-editor          → tb-snd-do(EDITORMANAGER, tbterm(close-editor))
```

### Compiler

```
       rec(module-manager >> compiler, parse)  →
                               tb-rec-msg(module-manager, compiler, tbterm(parse)) .
                               tb-snd-eval(COMPILER, tbterm(parse))
       parse-ok              → tb-rec-value(COMPILER, tbterm(parse-ok))
       parse-uptodate        → tb-rec-value(COMPILER, tbterm(parse-uptodate))
       parse-error           → tb-rec-value(COMPILER, tbterm(parse-error))
       rec(module-manager >> compiler, compile)   →
                               tb-rec-msg(module-manager, compiler, tbterm(compile)) .
                               tb-snd-do(COMPILER, tbterm(compile))
       compile-ok            → tb-rec-event(COMPILER, tbterm(compile-ok)) .
                               tb-snd-ack-event(COMPILER, tbterm(compile-ok))
       compile-uptodate      → tb-rec-event(COMPILER, tbterm(compile-uptodate)) .
                               tb-snd-ack-event(COMPILER, tbterm(compile-uptodate))
       compile-error         → tb-rec-event(COMPILER, tbterm(compile-error)) .
                               tb-snd-ack-event(COMPILER, tbterm(compile-error))
       rec(module-manager >> compiler, flatten)    →
                               tb-rec-msg(module-manager, compiler, tbterm(flatten)) .
       flatten-ok            → tb-rec-event(COMPILER, tbterm(flatten-ok)) .
                               tb-snd-ack-event(COMPILER, tbterm(flatten-ok))
       flatten-uptodate      → tb-rec-event(COMPILER, tbterm(flatten-uptodate)) .
                               tb-snd-ack-event(COMPILER, tbterm(flatten-uptodate))
       flatten-error         → tb-rec-event(COMPILER, tbterm(flatten-error)) .
                               tb-snd-ack-event(COMPILER, tbterm(flatten-error))
```

### ErrorViewer

```
       rec(compiler >> errorviewer, errors)   →
                               tb-rec-msg(compiler, errorviewer, tbterm(errors)) .
                               tb-snd-do(ERRORVIEWER, tbterm(errors))
       rec(animation-generator >> errorviewer, errors)   →
                               tb-rec-msg(animation-generator, errorviewer,
                                   tbterm(errors)) .
                               tb-snd-do(ERRORVIEWER, tbterm(errors))
```

### LibraryManager

```
       set-libraries         → tb-rec-event(MODULEMANAGER, tbterm(set-libraries)) .
                               tb-snd-ack-event(MODULEMANAGER, tbterm(set-libraries))
```

### Simulator

```
       rec(module-manager >> simulator, new-tilspecification)→
                               tb-rec-msg(module-manager, simulator,
                                   tbterm(new-tilspecification)) .
                               tb-snd-do(SIMULATOR, tbterm(new-tilspecification))
       rec(module-manager >> simulator, delete-tilspecification)   →
                               tb-rec-msg(module-manager, simulator,
                                   tbterm(delete-tilspecification)) .
                               tb-snd-do(SIMULATOR, tbterm(delete-tilspecification))
       rec(animation-generator >> simulator, new-animation)  →
```

```
                              tb-rec-msg(animation-generator, simulator,
                                  tbterm(new-animation)) .
                              tb-snd-do(SIMULATOR, tbterm(new-animation))
    rec(animation-generator >> simulator, animationgeneration-error) →
                              tb-rec-msg(animation-generator, simulator,
                                  tbterm(animationgeneration-error)) .
                              tb-snd-do(SIMULATOR, tbterm(animationgeneration-error))
```

**AnimationGenerator**

```
    rec(module-manager >> animation-generator, new-tilspecification) →
                              tb-rec-msg(module-manager, animation-generator,
                                  tbterm(new-tilspecification)) .
                              tb-snd-do(ANIMATIONGENERATOR,
                                  tbterm(new-tilspecification))
    rec(module-manager >> animation-generator, delete-tilspecification)   →
                              tb-rec-msg(module-manager, animation-generator,
                                  tbterm(delete-tilspecification)) .
                              tb-snd-do(ANIMATIONGENERATOR,
                                  tbterm(delete-tilspecification))
    new-animation         → tb-rec-event(ANIMATIONGENERATOR,
                                  tbterm(new-animation)) .
                              tb-snd-ack-event(ANIMATIONGENERATOR,
                                  tbterm(new-animation))
    animationgeneration-error →
                              tb-rec-event(ANIMATIONGENERATOR,
                                  tbterm(animationgeneration-error)) .
                              tb-snd-ack-event(ANIMATIONGENERATOR,
                                  tbterm(animationgeneration-error))
```

We rename all component modules and their main processes by putting a P in front of the original names, indicating a Process in the ToolBus, to distinguish them from the tools for which we prefix with a T. For possible adapters to be used with a tool we use an A as prefix.

*4.2 Constraining*

We constrain the ToolBus proceses obtained in the previous section with the specification of the tools. The specification of the tools is given in separate modules and each constraining of a ToolBus processes is done as shown below for the module manager

```
process module PT-ModuleManager
begin
   exports
   begin
      processes
         PT-ModuleManager
   end
   imports
      PModuleManager,
      TModuleManager
   definitions
      PT-ModuleManager = PModuleManager ∥ TModuleManager
end PT-ModuleManager
```

In the following we give the specification of tools.

**Function**

```
        TFunction =
           tooltb-snd-event(tbterm(quit)) .
           tooltb-rec-ack-event(tbterm(quit))
```

**Module Manager**

We decide that the module manager is responsible for the parsing of all modules, and that a compile request for a particular module is also a flatten request for this module. The compiler is responsible for compiling other modules this particular module depends on.

A compilation request for a module results in a series of messages on compilation results of the modules it depends on, ending in an error result or the result of the flattening of the module.

```
TModuleManager =
   (
      new-module
   + delete-module
   + tooltb-snd-event(tbterm(edit-module)) .
      tooltb-rec-ack-event(tbterm(edit-module))
   + tooltb-snd-event(tbterm(close-module)) .
      tooltb-rec-ack-event(tbterm(close-module))
   + tooltb-rec(tbterm(module-closed))
   + tooltb-rec(tbterm(module-written))
   + tooltb-snd-event(tbterm(parse)) .
      tooltb-rec-ack-event(tbterm(parse)) .
      (
         tooltb-rec(tbterm(parse-ok))
      + tooltb-rec(tbterm(parse-uptodate))
      + tooltb-rec(tbterm(parse-error))
      )
   + tooltb-snd-event(tbterm(compile)) .
      tooltb-rec-ack-event(tbterm(compile)) .
      (
         (
            tooltb-rec(tbterm(compile-ok))
         + tooltb-rec(tbterm(compile-uptodate))
         ) * (
            tooltb-rec(tbterm(compile-error))
         + tooltb-rec(tbterm(flatten-ok))
         + tooltb-rec(tbterm(flatten-uptodate))
         + tooltb-rec(tbterm(flatten-error))
         )
      )
   + tooltb-snd-event(tbterm(new-tilspecification)) .
      tooltb-rec-ack-event(tbterm(new-tilspecification))
   + tooltb-snd-event(tbterm(delete-tilspecification)) .
      tooltb-rec-ack-event(tbterm(delete-tilspecification))
   ) * delta
```

**Editor Manager**

For the editor manager we use recursion to keep track on the number of open editor sessions. This is necessary for internal action of the manager to act on open sessions.

```
TEditorManager = TEditorManager(nat(^0))
TEditorManager(n) =
      tooltb-rec(tbterm(start-editor)) .
      TEditorManager(succ(n))
   + [gt(n, nat(^0)) = true] → (
         tooltb-snd-event(tbterm(editor-close)) .
         tooltb-rec-ack-event(tbterm(editor-close)) .
         TEditorManager(pred(n))
      + tooltb-snd-event(tbterm(editor-write)) .
         tooltb-rec-ack-event(tbterm(editor-write)) .
         TEditorManager(n)
      )
   + tooltb-rec(tbterm(close-editor)) .
      TEditorManager(pred(n))
```

**Compiler**

On a compile request for a particular module the compiler has to compile all the modules this particular depends on. The compiler sends result messages for compiling of each module separately. Compilation of modules is stopped immediately on an error result. When there is no compilation error, the module of the request is flattened and a result message of the flattening is send.

Note that we expect separate request for parsing of each module on forehand. The compilation process sends a error result when it needs a module for which the parsing resulted in an error.

```
TCompiler =
   (
      tooltb-rec(tbterm(compile)) . (
         (
            compile-ok .
            tooltb-snd-event(tbterm(compile-ok)) .
            tooltb-rec-ack-event(tbterm(compile-ok))
         +  compile-uptodate .
            tooltb-snd-event(tbterm(compile-uptodate)) .
            tooltb-rec-ack-event(tbterm(compile-uptodate))
         ) * (
            compile-error .
            tooltb-snd-event(tbterm(compile-error)) .
            tooltb-rec-ack-event(tbterm(compile-error))
         +  flatten-ok .
            tooltb-snd-event(tbterm(flatten-ok)) .
            tooltb-rec-ack-event(tbterm(flatten-ok))
         +  flatten-uptodate .
            tooltb-snd-event(tbterm(flatten-uptodate)) .
            tooltb-rec-ack-event(tbterm(flatten-uptodate))
         +  flatten-error .
            tooltb-snd-event(tbterm(flatten-error)) .
            tooltb-rec-ack-event(tbterm(flatten-error))
         )
      )
   +  tooltb-rec(tbterm(parse)) . (
         parse-ok .
         tooltb-snd(tbterm(parse-ok))
      +  parse-uptodate .
         tooltb-snd(tbterm(parse-uptodate))
      +  parse-error .
         tooltb-snd(tbterm(parse-error))
      )
   ) * delta
```

**Error Viewer**

```
TErrorViewer =
   (
      tooltb-rec(tbterm(errors))
   ) * delta
```

**Library Manager**

```
TLibraryManager =
   (
      tooltb-snd-event(tbterm(set-libraries)) .
      tooltb-rec-ack-event(tbterm(set-libraries))
   ) * delta
```

**Simulator**

We use recursion to keep track on whether simulation is going on.

```
TSimulator = TSimulator(false)
TSimulator(simulating) =
      tooltb-rec(tbterm(new-tilspecification)) .
      TSimulator(simulating)
   +  tooltb-rec(tbterm(delete-tilspecification)) .
      TSimulator(simulating)
   +  tooltb-rec(tbterm(new-animation)) .
      TSimulator(simulating)
   +  tooltb-rec(tbterm(animationgeneration-error)) .
      TSimulator(simulating)
```

```
                +  [simulating = false]→ (
                      simulator-start .
                      TSimulator(true)
                   )
                +  [simulating = true]→ (
                      simulator-stop .
                      TSimulator(false)
                   +  simulator-quit .
                      TSimulator(false)
                   )
```

**Animation Generator**

```
           TAnimationGenerator =
             (
                tooltb-rec(tbterm(new-tilspecification))
             +  tooltb-rec(tbterm(delete-tilspecification))
             +  tooltb-snd-event(tbterm(new-animation)) .
                tooltb-rec-ack-event(tbterm(new-animation))
             +  tooltb-snd-event(tbterm(animationgeneration-error)) .
                tooltb-rec-ack-event(tbterm(animationgeneration-error))
             ) * delta
```

*4.3 The ToolBus Application*

We compose the system by importing the constrained ToolBus processes as instances of the NewTool module from the ToolBus library, and merging them into the system process.

```
       process module IDESystem
       begin
          exports
          begin
             processes
                IDESystem
          end
          imports
                ⋮
             NewTool {
                Tool bound by [
                   Tool → PT-ModuleManager
                ] to PT-ModuleManager
                renamed by [
                   TBProcess → ModuleManager
                ]
             },
                ⋮
          definitions
             IDESystem =
                   ModuleManager
                ‖  EditorManager
                ‖  Compiler
                ‖  ErrorViewer
                ‖  Simulator
                ‖  AnimationGenerator
       end IDESystem
```

And we put the system in the ToolBus application environment by importing the system as instance of the NewToolBus module from the ToolBus library.

```
       process module IDE
       begin
          imports
             NewToolBus {
                Application bound by [
                   Application → IDESystem
                ] to IDESystem
                renamed by [
```

```
                ToolBus → IDE
            ]
    }
end IDE
```

The resulting animation is shown in Figure 4.

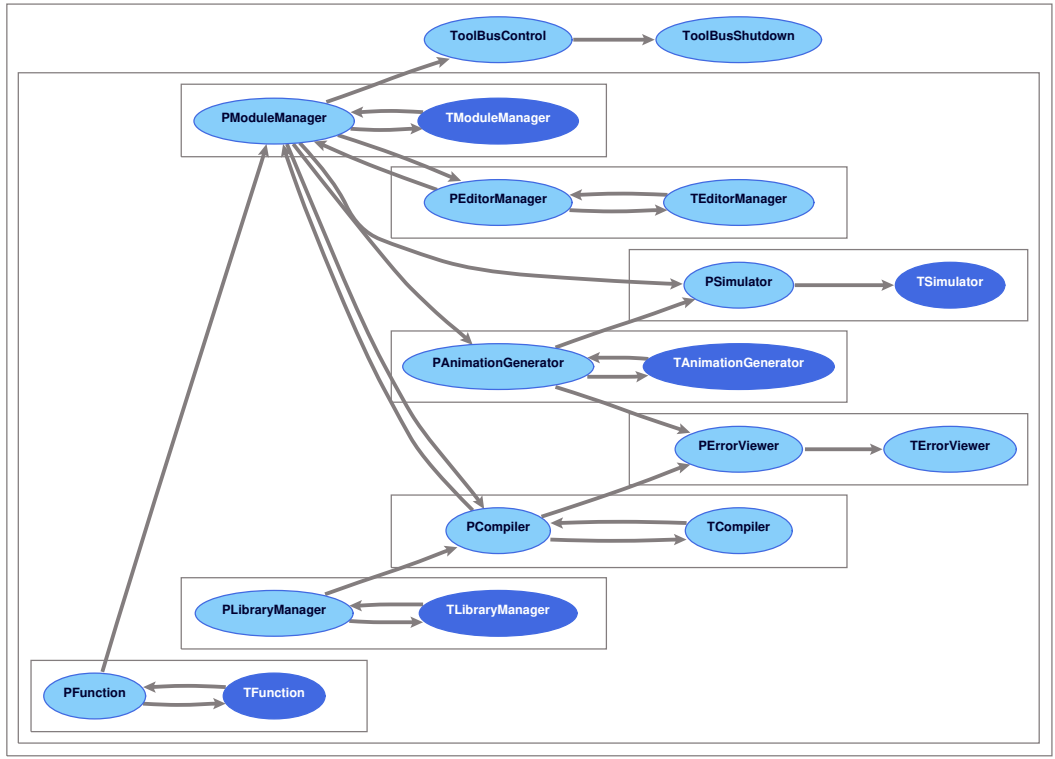

**Figure 4.** Animation of the IDE as ToolBus application

## 5. Implementation of the IDE

In the previous section we gave specifications of the tools which together make up the IDE. Although these specifications are rough, we consider them fine enough to proceed with the implementation of these tools.

*5.1 Implementation of the Tools*

**Function**

We implemented the Function component in the language Tcl/Tk [17]. It consist fo a single button for requesting to quit the IDE. We added another button to select a type of editor, since the implementation of the editor manager makes a choice of editor possible (see below). We updated the specifications to reflect this possibility.

**Module Manager**

The module manager controls the operation of the command issued by the user and gives information on the state of the partial compilation of the modules through a table. Tcl/tk is used as implementation language.

A PSF module must have a header and a trailer containing the module-name. This seems superfluous, since a file may only contain one module int he setting of the IDE. However, we have decided not alter the module structure. Instead, we generate the header and trailer whenever the user request a new module.

**Editor Manager**

An editor manager has not only as task to execute an editor on request, but also the managing of open sessions and marshalling interaction with the editors. Most development environment force an editor upon the user. The user may not be familiar with this editor and may even have to know several editors if working with different development environments. Ideally, the user may choose an editor with which a development environment should interact.

We have chosen to reuse the work of de Jong and Kooiker [12]. They implemented an editor manager which supports interactive editing with the popular editors GNU Emacs [18] and Vim[2] [16]. The manager is implemented as a ToolBus tool and can be integrated in the IDE without any modifications. It is implemented in the C programming language [13].

**Compiler**

The compiler acts as a controller for the parser, compiler, and flattener from the PSF Toolkit. It keeps track of imports by extracting imported modules from a parsed module. The imports are used to decide on the order of compilation steps of the (intermediate) modules. The compiler is implemented in Perl [19].

In the setting of the IDE, the parser may allow only one module per file and the name of the module and file must match. Instead of altering the parser to check on this, we implemented a separate check routine that the compiler invokes prior to the parser.

**Error Viewer**

We implemented the error viewer in Tcl/Tk. It consist of a display and a button to clear the display.

**Simulator**

The simulator is a wrapper for the simulator from the PSF Toolkit and is implemented in Tcl/Tk.

**Animation Generator**

The animation generator is a wrapper for the animation from the PSF Toolkit that provides control over the many command-line options. It is implemented in Tcl/Tk.

*5.2 ToolBus Script*

The ToolBus script for controlling the separate tools of the simulator can be derived from the ToolBus processes in the specification of the simulator as ToolBus application. This transformation is done by hand mainly because in the specification recursion is used to hold the state of a process and in a ToolBus script this has to be done with iteration and state variables. Also the data terms have to be refined to contain arguments necessary for identifying the module the message relates to.

*5.3 Aggregated GUI*

Except for the editor manager and compiler, each tool has its own graphical user interface (gui). Because several windows on the screen belonging to one application does not look very appealing, and opening several editor session makes this even worse, it is better to have one big gui for the IDE. Integrating editor sessions in this gui is not a good idea, since we make use of existing editors, of which the gui's do not fit in the gui of the IDE very well.

---

2. Vim is an improved version of vi, an editor distributed with most Unix-like operating systems.

In Tcl/Tk it is possible to indicate that a frame window is to serve as a container of another application and that a toplevel window is to be used as the child of such a container window. Following this scheme, we implemented a separate tool that does the layout of several container windows. This layout can be resized as a whole and some windows can be resized in relation to each other through the use of paned windows.[3] A user preferring a different layout can implement another version similar to this.

For a toplevel window to act as a child of a container window, it needs the identifier of the container window. The aggregated gui implementation has to communicate a window identifier to each child. The ToolBus script has been supplied with an initialization phase that receives all the identifiers of the container windows from the aggregated gui and distributes them over the tools. Each tool now first receives its parent identifier before doing anything else. The resulting gui is shown in Figure 5.

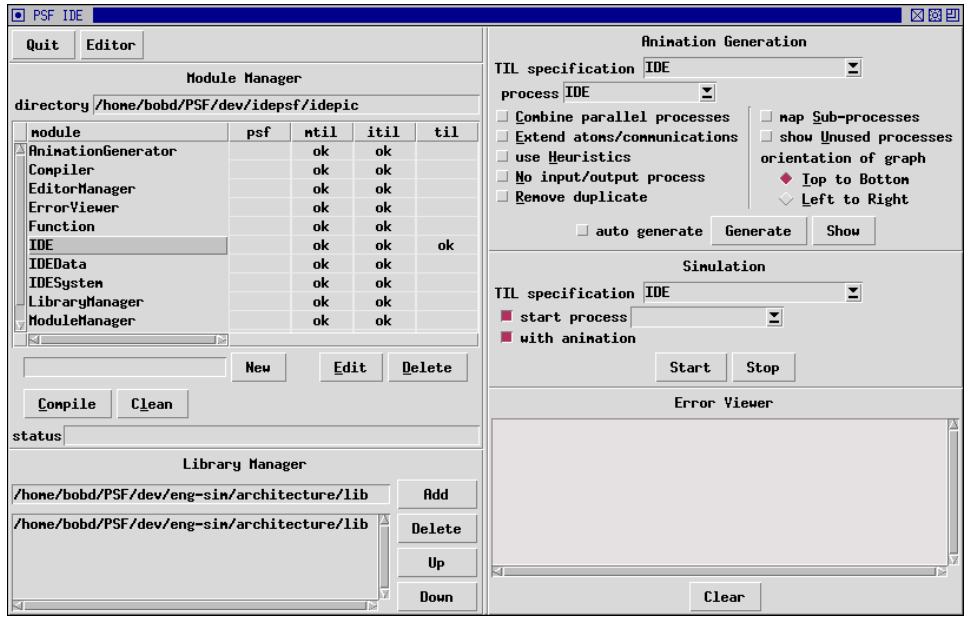

**Figure 5.** Aggregation of gui's

## 6. Conclusions

We engineered an IDE for PSF with the use of PSF, starting with a specification of the architecture. And by applying refining techniques, we obtained a specification of the IDE as a ToolBus application. We extracted a ToolBus script from the last specification and implemented the components as separate tools. To be integrated, the tools from the PSF Toolkit needed no modifications. The components are kept small and are easily replaceable due to the use of the ToolBus as coordination architecture.

The PSF Toolkit proofed to be very useful in the development process of the specifications, and we encountered no problems in the use of the PSF libraries for architecture specification and ToolBus specification. The development of the IDE has strengthen our conviction that process algebra should be used in the software engineering process and that PSF with the toolkit and the libraries is very suitable here.

**References**

[1] J.A. Bergstra and J.W. Klop, "Process algebra: specification and verification in bisimulation semantics," in *Math. & Comp. Sci. II*, ed. M. Hazewinkel, J.K. Lenstra, L.G.L.T. Meertens, CWI Monograph 4, pp. 61-94, North-Holland, Amsterdam, 1986.

---

3. A paned window consists of two horizontal or vertical panes separated by a movable sash, and each pane containing a window.

[2] J.A. Bergstra, J. Heering, and P. Klint (eds.), "The Algebraic Specification Formalism ASF," in *Algebraic Specification*, ACM Press Frontier Series, pp. 1-66, Addison-Wesley, 1989.

[3] J.A. Bergstra and P. Klint, "The discrete time ToolBus," *Science of Computer Programming*, vol. 31, no. 2-3, pp. 205-229, July 1998.

[4] J.A. Bergstra and C.A. Middelburg, "An Interface Group for Process Components," report PRG0707, Programming Research Group - University of Amsterdam, October 2007.

[5] B. Diertens, "Software (Re-)Engineering with PSF," report PRG0505, Programming Research Group - University of Amsterdam, October 2005.

[6] B. Diertens, "Software (Re-)Engineering with PSF II: from architecture to implementation," report PRG0609, Programming Research Group - University of Amsterdam, November 2006.

[7] B. Diertens, "New Features in PSF I - Interrupts, Disrupts, and Priorities," report P9417, Programming Research Group - University of Amsterdam, June 1994.

[8] B. Diertens and A. Ponse, "New Features in PSF II - Iteration and Nesting," report P9425, Programming Research Group - University of Amsterdam, October 1994.

[9] B. Diertens, "Simulation and Animation of Process Algebra Specifications," report P9713, Programming Research Group - University of Amsterdam, September 1997.

[10] B. Diertens, "Generation of Animations for Simulation of Process Algebra Specifications," report P2003, Programming Research Group - University of Amsterdam, September 2000.

[11] E.R. Gansner and S.C. North, "An Open Graph Visualization System and its Applications to Software Engineering," *Software -- Practice and Experience*, vol. 30, no. 11, pp. 1203-1233, 2000.

[12] H.A. de Jong and A.T. Kooiker, *My Favourite Editor Anywhere,* Lecture Notes in Computer Science, 3475, pp. 122-131, Springer-Verlag, 2005.

[13] B.W. Kernighan and D.M. Ritchie, *The C Programming Language,* second edition, Prentice Hall, Englewood Cliffs, New Jersey, 1988.

[14] S. Mauw and G.J. Veltink, "A Process Specification Formalism," in *Fundamenta Informaticae XIII (1990)*, pp. 85-139, IOS Press, 1990.

[15] S. Mauw and G.J. Veltink (eds.), *Algebraic Specification of Communication Protocols,* Cambridge Tracts in Theoretical Computer Science 36, Cambridge University Press, 1993.

[16] B. Molenaar, *Vim the editor*, website: www.vim.org.

[17] J.K. Ousterhout, *Tcl and the Tk Toolkit,* Addison-Wesley, 1994.

[18] R.M. Stallman, "Emacs the Extensible, Customizable Self-Documenting Display Editor," *Proceedings of the ACM SIGPLAN SIGOA symposium on Text manipulation*, pp. 147-156, 1981.

[19] L. Wall, T. Christiansen, and R.L. Schwartz, *Programming Perl,* O'Reilly & Associates, Inc., 1996.